\documentclass[ amsmath,amssymb,nofootinbib, aps, pra, reprint, longbibliography]{revtex4-1}
\usepackage{dcolumn}
\usepackage{graphicx}
\usepackage{amsfonts} 
\usepackage{epstopdf}
\usepackage{color}
\usepackage{ulem}
\usepackage{here}
\usepackage[dvipsnames]{xcolor}
\usepackage{hyperref}
\usepackage{pifont}

\AtBeginDocument{
    \newwrite\bibnotes
    \def\bibnotesext{Notes.bib}
    \immediate\openout\bibnotes=\jobname\bibnotesext
    \immediate\write\bibnotes{@CONTROL{REVTEX41Control}}
    \immediate\write\bibnotes{@CONTROL{%
    apsrev41Control,author="08",editor="1",pages="1",title="0",year="1"}}
     \if@filesw
     \immediate\write\@auxout{\string\citation{apsrev41Control}}%
    \fi
}

\begin{document} 
\title{Optimizing optical potentials with physics-inspired learning algorithms}

\author{M. Calzavara$^{1,4,*}$, Y. Kuriatnikov$^{2,*}$, A. Deutschmann-Olek$^{3}$, F. Motzoi$^{1}$, S. Erne$^{2}$, A. Kugi$^{3}$, T. Calarco$^{1,4}$,  J. Schmiedmayer$^{2}$, M. Pr\"ufer$^{2,\S}$}
 \affiliation{$^1$Forschungszentrum J\"ulich GmbH, Peter Gr\"unberg Institute, Quantum Control (PGI-8), 52425 J\"ulich, Germany\\
 $^2$Vienna Center for Quantum Science and Technology, Atominstitut, TU Wien, Stadionallee 2, 1020 Vienna, Austria\\
 $^3$Automation and Control Institute, TU Wien, Gu{\ss}hausstra{\ss}e 27-29, 1040 Vienna, Austria\\
 $^4$Institute for Theoretical Physics, Universit\"at zu K\"oln, 50937 Cologne, Germany}
 
\begin{abstract}

We present our experimental and theoretical framework which combines a broadband superluminescent diode (SLD/SLED) with fast learning algorithms to provide speed and accuracy improvements for the optimization of 1D optical dipole potentials, here generated with a Digital Micromirror Device (DMD). To characterize the setup and potential speckle patterns arising from coherence, we compare the superluminescent diode to a single-mode laser by investigating interference properties. We employ Machine Learning (ML) tools to train a physics-inspired model acting as a digital twin of the optical system predicting the behavior of the optical apparatus including all its imperfections. Implementing an iterative algorithm based on Iterative Learning Control (ILC) we optimize optical potentials an order of magnitude faster than heuristic optimization methods. We compare iterative model-based “offline”  optimization and experimental feedback-based “online” optimization. Our methods provide a route to fast optimization of optical potentials which is relevant for the dynamical manipulation of ultracold gases.
\end{abstract}
\maketitle
\def\thefootnote{*}\footnotetext{These authors contributed equally to this work}
\def\thefootnote{ \S}\footnotetext{maximilian.pruefer@tuwien.ac.at}

\section{Introduction}

The precise control and manipulation of light fields are required for many diverse areas of research ranging from microscopy \cite{PhysRevLett.120.193901} to quantum simulators \cite{Gross:17}.
In particular, optical beam shaping constitutes a common task, for which wavefront manipulating devices, such as Spatial Light Modulators (SLM) or Digital Micromirror Devices (DMD), are especially suited.
The beam shaping is important for experiments with ultracold gases, where optical dipole potentials have proven to be a versatile tool to provide the demanded level of control.
In combination with a DMD, almost arbitrary shaping of the optical potential is possible, both in 1D \cite{ Henderson_2009,  Gauthier:16,  PhysRevA.95.013632, Tajik:19} and 2D \cite{ Liang:10, PhysRevLett.110.200406, navon_smith_hadzibabic_2021, GAUTHIER20211, Zupancic:16} settings. These potentials can, as an example, be used for generating homogeneous box potentials in ultracold gases experiments \cite{navon_smith_hadzibabic_2021}.
In addition to static potentials, dynamical perturbations and time-averaged potentials can also be created, by projecting sequences of patterns onto the DMD \cite{Gauthier:16,PhysRevResearch.3.033241}.

There exist two main approaches to the optimization of optical potentials: precalculating DMD patterns based on physical assumptions \cite{Floyd:76,Dorrer:07} and models \cite{Liang:10} or iteratively updating DMD patterns based on experimental feedback \cite{Liang:10, Tajik:19}.
The first approach avoids the need for feedback measurement but is limited by model precision and thus requires detailed system characterization, while the second gives the most accurate results but typically requires a large number of experimental iterations.
Here we implement a purely data-driven approach that combines different learning algorithms to get the best of both worlds. 

Using a digital twin of the system makes it possible to shape different types of target potentials without the need for experimental feedback. Yet because of speckles caused by imperfections, a model featuring just a few known experimental parameters (parametric model) can only predict its behavior up to limited precision and might not always be suitable for precise potential shaping.
In some cases, though, such as in “clean" systems with pin-hole filtering, simulations combined with input beam characterization deliver very low errors on precalculated DMD patterns \cite{Liang:10}.
We improve the prediction performance compared to parametric models by employing data-driven learning techniques. 
Learning methods are already used for estimating the transfer matrix of complex optical systems \cite{caramazza_moran_murray-smith_faccio_2019} and provide good results. As their main disadvantage, they generally require a large amount of data for sufficient training. In our approach, we develop a physics-inspired model that requires a smaller amount of training data and thus saves experimental time.

Despite any improvement in model precision, the effect of residual error sources can only be mitigated by using experimental feedback \cite{Liang:10, Tajik:19}.
Therefore, we introduce a feedback optimization method based on Iterative Learning Control (ILC) \cite{ILCrev2006,Deutschmann-Olek:22}. This method is directly applicable to various types of experiments with wavefront manipulating devices. Since system knowledge is directly employed in the update law that adjusts the DMD setting based on feedback, the number of required experimental iterations is significantly reduced compared to heuristic methods \cite{Tajik:19}.

Because of the learning-based nature of both the ML model and ILC method, they benefit from high predictability of the system behavior, which can be achieved by using light sources with low coherence.
White light was used for Bose-Einstein condensates (BEC) trapping in order to minimize the impact of speckles on density fluctuations \cite{Sackett_2003},
and the advantages of using a superluminescent diode (SLD) in comparison with a single-frequency laser were shown also in combination with a DMD \cite{PhysRevResearch.3.033241}. 
Indeed, we find that using the SLD improves model prediction results, while the feedback-based methods give errors comparable to measurement errors for both light sources.

In perspective, the ability to efficiently shape a large number of potentials using a DMD will provide a high level of control for the dynamical manipulation of quantum gases.
The realization of non-harmonic optimal protocols \cite{Campo:12} and quantum thermal machines \cite{Gluza:21} are but two examples among countless possible applications.

The paper is organized as follows: in Sec.\ref{sec:Experiment} we describe our experimental setup and test the coherence properties of the light sources by measuring interferences of a SLD in comparison with a single-frequency laser; in Sec.\ref{sec:Model} we introduce the physics-inspired learning model used for representing the optical system including its imperfections; in Sec.\ref{sec:Optimization} we use iterative learning control algorithms for optimizing optical dipole potentials; in Sec.\ref{sec:Conclutions} we summarize our results. 

%%%%%%%%%%%%%%%%%%%%%%%%%%%%%%%%%%
\section{Experimental setup}\label{sec:Experiment}
%%%%%%%%%%%%%%%%%%%%%%%%%%%%%%%%%%
\begin{figure}[t]
	\centering
	\includegraphics[width=3.4in]{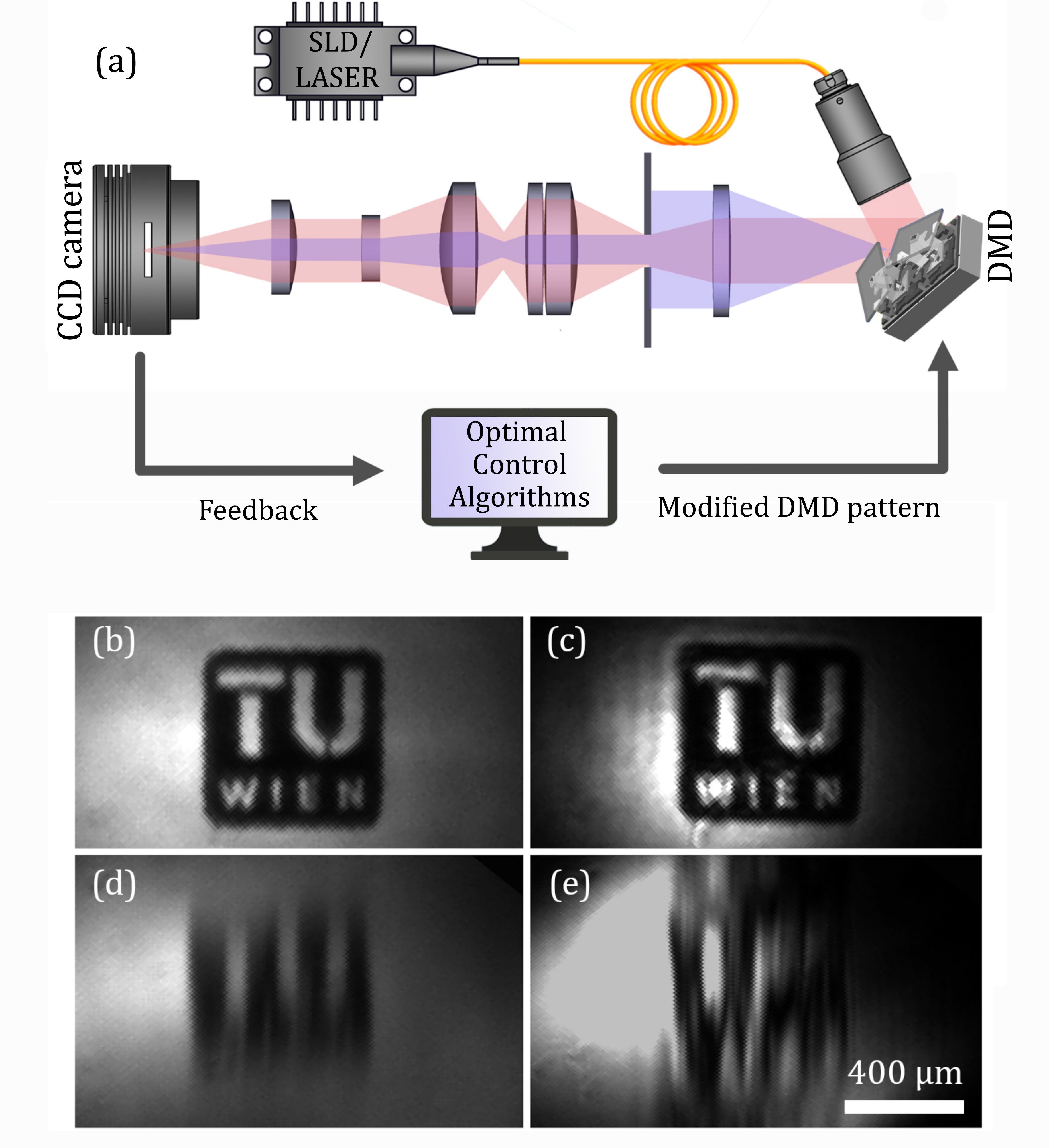}
	\caption{\textbf{{(a)}} Schematic of the experimental setup. The light source is an interchangeable SLD/laser connected with a fiber to the collimator and projected onto the DMD. The DMD is an array of individually controllable micromirrors which allows for projecting arbitrary patterns. The system is shown as simplified schematic omitting mirrors.  The shown lens system constitutes an imaging of the DMD pixels into the CCD camera (image) plane with a demagnification of 17.5 and minimizing chromatic aberrations. The slit is placed in an effective Fourier plane of the imaging system. The red path corresponds to the beam and the blue to a point source in the DMD plane. \textbf{{(b)-(e)}} Images taken with the CCD camera in the image plane of the DMD with the same pattern projected for \textbf{{(b)}} SLD without slit \textbf{{(c)}} laser without slit \textbf{{(d)}} SLD with slit \textbf{{(e)}} laser with slit.}
    \label{fig:setup}
\end{figure}
\begin{figure*}[t]
	\includegraphics[width = \textwidth]{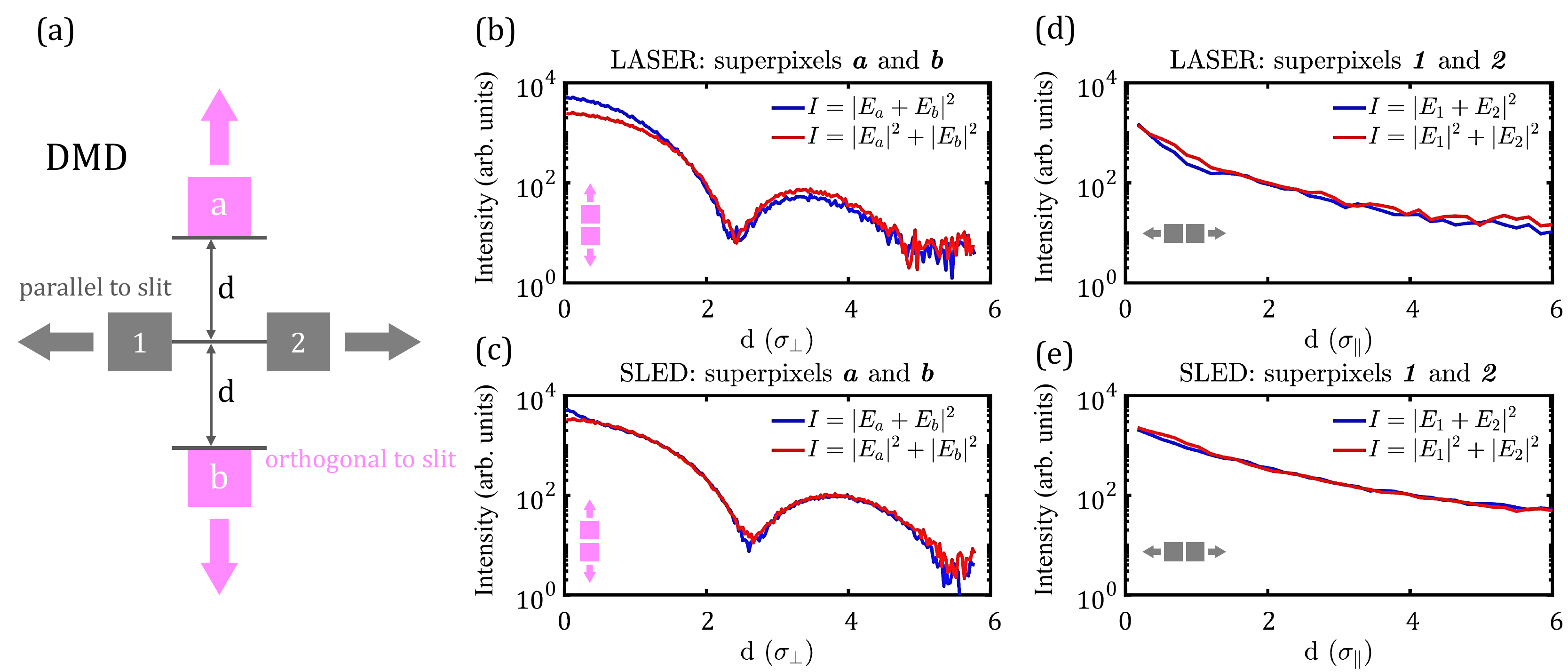}
   	\caption{\textbf{{(a)}} Schematics of test measurement of coherence properties. Two pairs of 3x3 DMD superpixels \textbf{{a/b}} and \textbf{{1/2}} are translated in a direction orthogonal and parallel to the slit, respectively. The resolution of the optical system is $\sigma_\parallel=3.3\,\mu$m in the longitudinal (parallel to slit)  and $\sigma_\perp=25\,\mu$m in the transverse (orthogonal to slit) direction, with respect to the slit orientation. \textbf{{(b)-(e)}} Intensity measured between two superpixels as a function of distance $d$ between them; the blue solid line shows the intensity measured for both superpixels simultaneously “on” (electric fields from superpixels contribute in measured intensity as $|E_{a/1} + E_{b/2} |^2$), the red solid line is the sum of intensities measured with only one of two superpixels “on” (contribution is independent $|E_{a/1}|^2 + |E_{b/2} |^2$). The distance $d$ is given in units of $\sigma_\parallel$ and $\sigma_\perp$. When the laser-illuminated superpixels are moved orthogonal to the slit \textbf{{(b)}}, positive interference is visible in the $0^{th}$ slit diffraction maximum and negative interference in the $1^{st}$.  When SLD-illuminated superpixels are moved orthogonal to the slit \textbf{{(c)}} a slight interference effect is visible only in the $0^{th}$ slit diffraction maximum for very small distances between the superpixels. Panel \textbf{{(d)}} shows laser-illuminated superpixels moving parallel to the slit, and panel \textbf{{(e)}} shows SLD-illuminated superpixels moving parallel to the slit. There are no noticeable interference effects in the longitudinal direction on the scales larger than the camera pixel size for both light sources. }
    \label{fig:laser_sled_slit}
\end{figure*}

In this section, we describe our experimental setup and compare the optical coherence properties of a single-frequency laser and broadband SLD. The optical apparatus was designed and optimized for manipulating 1D optical dipole potentials in our atom chip experiment \cite{Manz:10}. The simplified optical setup is shown in Fig.\ref{fig:setup} (a). In the setup, the light source is easily interchangeable: we here use  {either} a superluminescent 110 mW fiber-coupled diode with a spectral width of 12.7 nm and a central wavelength of 837 nm  or a single-frequency 780\,nm laser.  The fiber is connected to a collimator and illuminates the DMD with collimated light. The DMD is placed in the focus of the first lens. The first two lenses together form a 4-f telescope. The focal point between the two lenses is a Fourier plane with respect to the DMD's image plane. The slit, which is adjustable in the transverse direction and is parallel to the observable 1D optical potential, is placed in the Fourier plane and closed to 0.625 mm. Being placed in the Fourier plane, the slit acts as a low-pass filter for the transverse spatial frequencies of the light field. This means it cuts off high-frequency $k$-modes of the DMD pattern effectively leading to a lowering of the resolution. The other 4 lenses form an objective designed to correct chromatic aberrations for the broadband SLD. The system acts with a demagnification of 17.5 and resolution $\sigma_\parallel=3.3\,\mu$m in the direction parallel to the slit (the resolution is given by our atom chip experiment {\cite{Manz:10}} for which the optical apparatus was designed and optimized) and $\sigma_\perp=25$ $\mu m$ in the transverse direction (orthogonal to the slit).
In the end, an image of the optical potential is acquired with a CCD camera sensor with 2.4 $\mu m$ pixel size. 

We use a $10.8$ $\mu m$ pitch near-infrared DMD, which is a 2D array of 1280x800 micromirrors. In the image plane (on the CCD camera) 5 DMD pixels correspond to $\sigma_\parallel$ in the longitudinal direction and 40 DMD pixels correspond to $\sigma_\perp$ in the transverse direction. Since many pixels in the transverse direction contribute to the local intensity (close to the center pixels contribute almost equally to the optical potential while far off-center pixels are contributing less), we can perform very smooth gray scaling, which is very important for high-precision optimization of 1D optical dipole potentials via DMD.
Yet, while a narrower slit allows for better gray scaling and with that lower discretization error and higher accuracy, it leads to  significantly reduced light intensity.  

The main advantage of using the SLD compared to the single-frequency laser as a light source is its reduced temporal coherence \cite{Hitzenberger:99, redding_choma_cao_2012} which is why random interferences (speckles) are reduced  (see Fig.\ref{fig:setup} (b,c)).
Here, we characterize coherence effects for both light sources as we want to understand whether we can consider the individual pixels as coherent or incoherent sources.

We prepare two series of patterns, each of them containing two superpixels moving away from each other (by superpixel, we understand a square pattern of 3x3 micromirrors). In the image plane on the CCD camera, the size of the superpixel is below the resolution of the optical apparatus and the superpixel can be considered as a point source, at the same time minimizing diffraction from the edges of individual DMD mirrors. We measure the intensity in the center between two superpixels. The imaged intensity is in general given by the squared sum of the electric fields of the individual pixels. In the incoherent case, it reduces to the sum of intensities from the individual pixels such that the system behaves linearly with respect to intensity.

 First, the pair of superpixels is moving orthogonal to the slit orientation (transverse direction), where the slit is closed to 0.625 mm and resolution is lowered to $\sigma_\perp$=25 $\mu m$. For both SLD and laser (see Fig.\ref{fig:laser_sled_slit} (b,c)), the slit diffraction maxima are clearly distinguishable. For the laser, positive interference can be observed in the $0^{th}$ maximum and negative in the $1^{st}$ one. For the SLD, interference is only observable on a very small scale (below $\sigma_\perp$).
Second, the superpixels are moving in the longitudinal direction, where the slit is fully opened to 13 mm and the resolution $\sigma_\parallel$=3.3 $\mu m$ is very close to the camera pixel size 2.4 $\mu m$. We observe only small interference effects for the laser, most likely because they are hardly accessible on this scale.

In conclusion, we observe that in both cases the SLD shows linear behavior, that is the total intensity is given by the incoherent sum of the intensities of each point source.
This showcases the advantages of the reduced coherence of the SLD, which we anticipate to decrease the amount of speckles and therefore to increase the predictability of the system.

%%%%%%%%%%%%%%%%%%%%%%%%%%%%%%%%%%
\section{Physics-inspired learning model}\label{sec:Model}
%%%%%%%%%%%%%%%%%%%%%%%%%%%%%%%%%%
\begin{figure*}[t]
	
	\includegraphics[width = \textwidth]{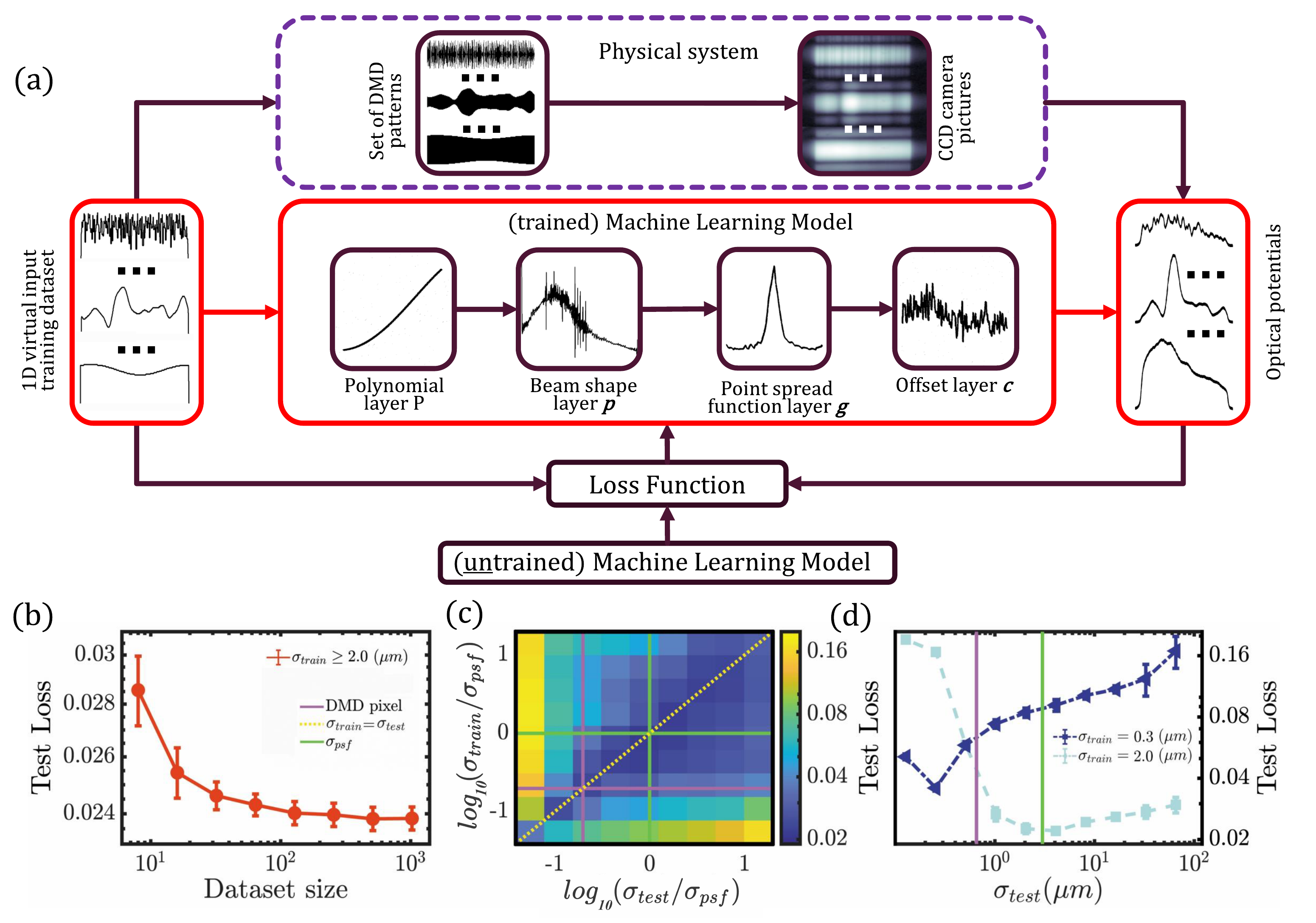}
	\caption{\textbf{{(a)}} Training of the physics-inspired model. The training dataset is composed of 1D virtual inputs $\boldsymbol{\nu}$ which are transformed into 2D DMD patterns (according to Eq. (\ref{eq:VRTtoDMD})) and projected in the image plane. We read out a pixel row from the CCD camera to infer the 1D optical potential generated by the system. After data acquisition, the 4-layer model is trained by minimizing the loss function. The polynomial layer represents the nonlinear relation between virtual input and local grayscale, while the other three layers take into account physical properties, i.e. the beam shape, the effective point spread function, and the background. The trained model (red solid segment) is a digital twin of the physical system which converts 1D virtual input to predicted optical potential. All the signals and layers shown in the figure are actual experimental data.  \textbf{{(b)}} Test loss dependence on the size of the training dataset. The loss converges to $2.4\%$ for a dataset size of $\sim10^2$, indicating that the model does not exhibit overfitting even for small training datasets. \textbf{{(c)}} Test loss dependence on the spatial frequency content of training and test datasets (see main text for details). \textbf{{(d)}} Two cuts from figure (c) for $\sigma_{train} = 0.3 $ $\mu m$ and $\sigma_{train} = 2.0$  $\mu m$.}
    \label{fig:ML}
\end{figure*}

\begin{figure*}[t]
	
	\includegraphics[width = \textwidth]{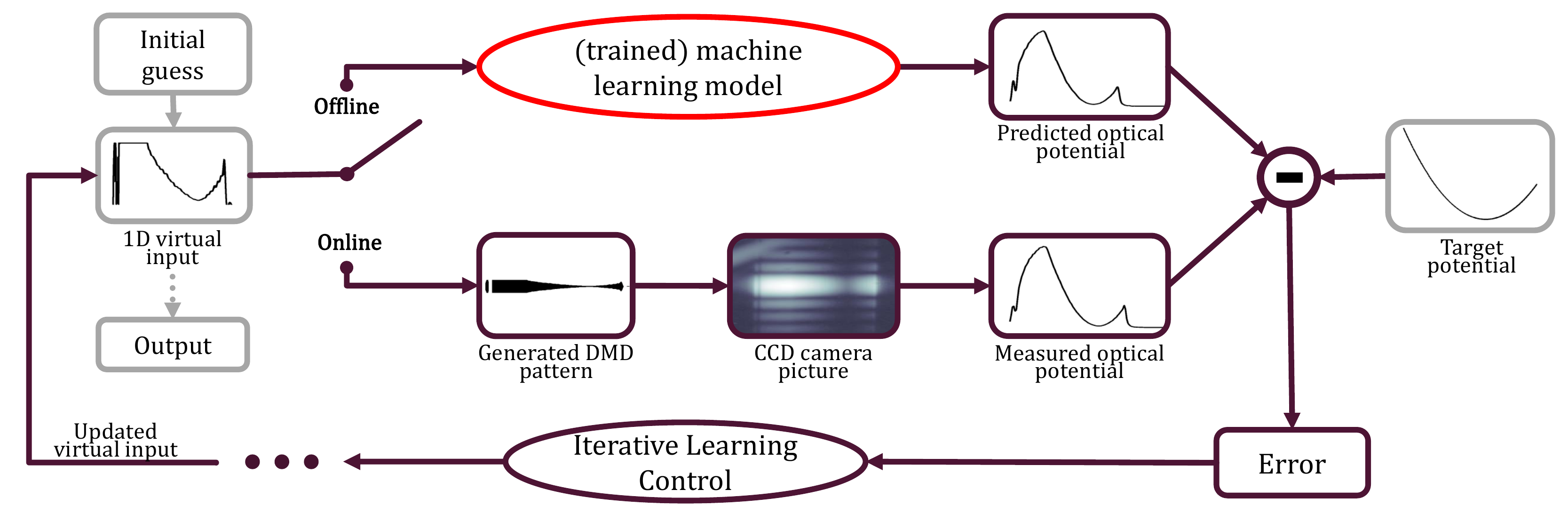}
		\caption{Scheme for optimizing optical potentials using Iterative Learning Control. The iterative scheme is initialized by choosing an initial guess and the target potential. We use two methods that differ in the feedback source: offline and online. The offline method uses the trained ML model to predict the optical potential. The online method projects the pattern on the DMD and measures the optical potential. Then the predicted (measured) potential is subtracted from the target to get the error that ILC uses to update the 1D virtual input. The procedure is repeated iteratively until the error converges or matches conditions for loop break. For the best performance, the output of the offline optimization can be used as the initial guess for online optimization. The result of the optimization loop is an optimized DMD pattern. All the signals in the figure are actual experimental data. }
    \label{fig:ML_ILC}
\end{figure*}

In this section, we describe how we use Machine Learning (ML) tools and experimental data to obtain a digital twin of the system. It should represent precisely the experimental system, while at the same time having a small number of
parameters to reduce the required number of experimental measurements. In the following, we outline our approach based on a physics-inspired model.

To formulate our problem, we treat the DMD as a 2D array to which we associate a binary configuration matrix $u_{ij}$.
For each pixel, a value of 1 corresponds to the mirror position that reflects light to the optical system, while 0 corresponds to the mirror position that reflects light away from the system.
The potential $\boldsymbol{V}$ is a vector obtained by selecting one row of camera pixels in the imaged output.
We refer to the coordinate along this row (and therefore, parallel to the slit) as $z$.
We then cast model training in the language of regression \cite{Vapnik2006EstimationOD}: given a set of
data points composed by $K$ couples $(u^{(k)}_{ij}, V^{(k)}_i)$,
and a family of functions $\mathcal{M}_{\boldsymbol{\alpha}}$,
parametrized by $\boldsymbol{\alpha}=[ \alpha_1, ... ,\alpha_N]$, we find the values of $\boldsymbol{\alpha}$
that minimize a loss function $L(u^{(k)}_{ij}, V^{(k)}_i, \boldsymbol{\alpha})$ defined below.

Regression problems are known to be affected by the bias-variance tradeoff \cite{Geman:92}.
This implies that large models, having $\dim(\boldsymbol{\alpha})>>K\dim(\boldsymbol{V})$, tend to approximate well the training data (low bias)
but fail to accurately generalize the prediction on test data (high variance).
This phenomenon is known as overfitting, and it puts a challenging limitation
to our program. 
In order to alleviate this problem, our approach consists in developing a physics-inspired model (see Fig.\ref{fig:ML} (a)) based on the knowledge we have about optical systems.
This way, we can reduce the number of its coefficients to the minimum
necessary to represent the system precisely enough, while avoiding overfitting.

First, we select an area of interest (AOI) on the DMD where the patterns will be located, based on the beam size
and position. All pixels outside the AOI are turned off. The AOI has a size of $(N_{row},N_{col})$, and its columns are orthogonal with respect to the slit.
 Then, we employ a dimensional reduction technique which we refer to as the “virtual input” 
(cf. \cite{Deutschmann-Olek:22}).
The virtual input measures the relative optical intensity induced by the pixels in the corresponding column.
Since the narrow slit 
averages along the transverse direction, we can use the pixels on each
column to create a set of gray scales at each point along the longitudinal direction $z$.
In the limit of a very narrow slit, all the pixels in a column contribute almost
equally to the final intensity, so we can define the virtual input $\boldsymbol{\nu}$ 
as the vector:
\begin{equation}
    \nu_j = \frac{1}{N_{row}}\sum_i u_{ij}\,.
    \label{eq:DMDtoVRT}
\end{equation}
This quantity represents the fraction of pixels that are turned on in each column,
with respect to the total number of rows inside the AOI.

In the case of non-zero slit widths, the mapping between a virtual input according to Eq. \eqref{eq:DMDtoVRT} and the actual relative optical intensity due to the column pattern
is non-linear and depends on the beam and on the transverse shape of the
point spread function (PSF). Therefore, we have to take this effect into account when we design our model.

Once $u_{ij}$ is transformed into $\nu_j$ some information is lost
since different binary matrices map onto the same virtual input.
To invert this mapping we have to choose a subset of binary matrices 
over which the mapping is one-to-one. 
We define the inverse map by turning on the pixels 
one by one on alternating rows above and below the central row of the AOI
$i_c$, according to the order $i_c, i_c+1, i_c-1, i_c+2, i_c-2,...$.
More compactly:
\begin{equation}
    u_{ij} = \theta\left(\frac{N_{row} \nu_j}{2}  - \lvert i  - i_c - \frac{1}{4}\rvert \right)\,,
    \label{eq:VRTtoDMD}
\end{equation}
where $\theta(x)$ is the Heaviside step function.

Using the virtual input and restricting the DMD configurations to the set described by Eq.\eqref{eq:VRTtoDMD} reduces the dimension of the input and output space, thus simplifying the learning problem. On the other hand, the restriction of the configurations might potentially introduce an artificial limitation to the potentials that can be realized.
In practice, we find the subset of binary matrices defined by this mapping to be wide enough so that it can be used for shaping arbitrary potentials.

For the task of learning the relation between the 1D input $\nu$ and the 1D potential $\boldsymbol{V}$ we propose the model $\boldsymbol{V} = \mathcal{M}_{\boldsymbol{\alpha}}(\boldsymbol{\nu})$ with (learnable) parameters $\boldsymbol{\alpha}$. This model is depicted in Fig.\ref{fig:ML} (a) and explicitly reads:
\begin{equation}
V_i = \left\lvert \sum_{j=-M}^{M} g_j P(\nu_{i-j}, q_1, ..., q_{N_P})p_{i-j} \right\lvert + c_i
\label{eq:ML_model}
\end{equation}
where $P(x,q_1,...,q_{N_P}) = \sum_{n=1}^{N_P} q_n x^n$ is a polynomial of degree $N_P=5$ with no constant term, and the vector of parameters $\boldsymbol{\alpha}$ is the concatenation of $\boldsymbol{g}$,$\boldsymbol{q}$,$\boldsymbol{p}$ and $\boldsymbol{c}$ with $\dim(\boldsymbol{\alpha}) \sim 1400$.
The polynomial function $P$ represents in an abstract way the non-linear relation between 
virtual input and resulting local relative intensity as discussed above.
The remaining parameters are chosen to mimic features of the physical system: the convolutional kernel $\boldsymbol{g}$ of size $N_g = 2M+1 \sim 71$ plays the role of the longitudinal PSF, the position-dependent term $\boldsymbol{p}$, of size $N_V \sim 650$, mimics the inhomogeneous light beam and the offset term $\boldsymbol{c}$, also of size $N_V$, gives the background.
In the convolution sum we employ zero padding, which means that the summand is zero if $i-j$ exceeds the index range $[1, N_V]$.

For the training, we choose the loss function as:
\begin{equation}
    L(\boldsymbol{\nu}^{(k)}, \mathbf{V}^{(k)}, \boldsymbol{\alpha}) = \sum_{k=1}^K \frac{\sum_{i=1}^{N_V} | [\mathcal{M}_{\boldsymbol{\alpha}}(\boldsymbol{\nu}^{(k)})]_i - V^{(k)}_i |}{\sum_{i=1}^{N_V} V^{(k)}_i}\,,
\end{equation}
which is the mean absolute error between prediction and measurement, normalized by the average potential itself.  This way, potentials with different average values will contribute equally during the training. 

To test the performance of the ML model, we create a data set of 10,000 random virtual inputs by sampling a white noise probability distribution. We filter the sample with Gaussian filters with 10 values of $\sigma_{data}$ ranging from
$0.2$ to $100$ DMD pixels (which corresponds to the range $[0.1,65] \mu$m mapped to the image plane). This way, the dataset was subdivided into 10 subsets with varying upper spatial cutoff frequency.
For each virtual input, the corresponding potential was measured and stored (along with the corresponding input).

We tested the dependence of the test loss on the training dataset size (see Fig.\ref{fig:ML} (b)). Both the training and test datasets were assembled by mixing 
the subsets with $\sigma_{data} \geq 2.0\ \mu$m.
The model is then trained on data chunks
of increasing size $K_{train}$, 
while keeping the test dataset size fixed to $K_{test}=300$. This sequence is repeated 4 times,
choosing new data sets at each time to compute the standard deviation of the test loss.
Even with $K_{train}=8$, the test loss is already below $3\%$ and the improvement for $K_{train} \sim 10^2$ is around half a percent.
Adding even more data points does not appreciably improve the prediction quality.
This result is particularly important since we aim for methods that are readily applicable to atomic density data measurement. 
In a typical experiment with trapped ultracold atomic gases,
taking more than $K \sim \mathcal{O}(10^2)$ atom densities pictures (with averaging over a few shots each) for potential optimization is a time-costly procedure.
The fact that the model we developed can be trained with less than 100 potentials means that it is in principle possible
to employ a data set obtained by atom-density estimation of the potential \cite{Tajik:19}.

To further analyze the learning process, we tested the dependence of the test loss on the cutoff frequency (see Fig.\ref{fig:ML} (c,d)) of both the training and test data sets. The frequency subsets are not mixed. At each time, we choose independently the cutoff frequency of the training $\sigma_{train}$ and test $\sigma_{test}$ datasets. 
The dataset sizes for training $K_{train} = 160$ and for testing $K_{test} = 39$ are fixed, and this sequence is repeated 5 times.
Fig.\ref{fig:ML} (c) shows qualitatively the test loss as a function of $\sigma_{train}$ and $\sigma_{test}$. The best results
are obtained close to the diagonal, that is, where $\sigma_{train} \sim \sigma_{test}$, so that training and test data look most similar.
The test loss becomes worse when $\sigma_{train}$ or $\sigma_{test}$ are around or below the DMD pixel size (magenta line on the plot),
indicating a possible mismatch between the behavior of the system and the ML model on the scale of the DMD pixel.
For a more quantitative interpretation, we show two different regimes in Fig.\ref{fig:ML} (d), where a representative scenario of the training on low frequencies (dark blue)
is compared to another curve representing the high-frequency case. In the first case, the test loss exhibits a low plateau above the DMD pixel size
and an abrupt increase below, corresponding to a breakdown of the low-frequency trained model in the high-frequency regime. 
The second curve instead shows how including the high frequencies during the training does not solve the problem, as now the prediction 
performance severely deteriorates along the whole frequency range.

%%%%%%%%%%%%%%%%%%%%%%%%%%%%%%%%%%%%%
\section{Potential optimization}\label{sec:Optimization}
%%%%%%%%%%%%%%%%%%%%%%%%%%%%%%%%%%%%%

\begin{figure*}[t]
	\centering
	\includegraphics[width=\textwidth]{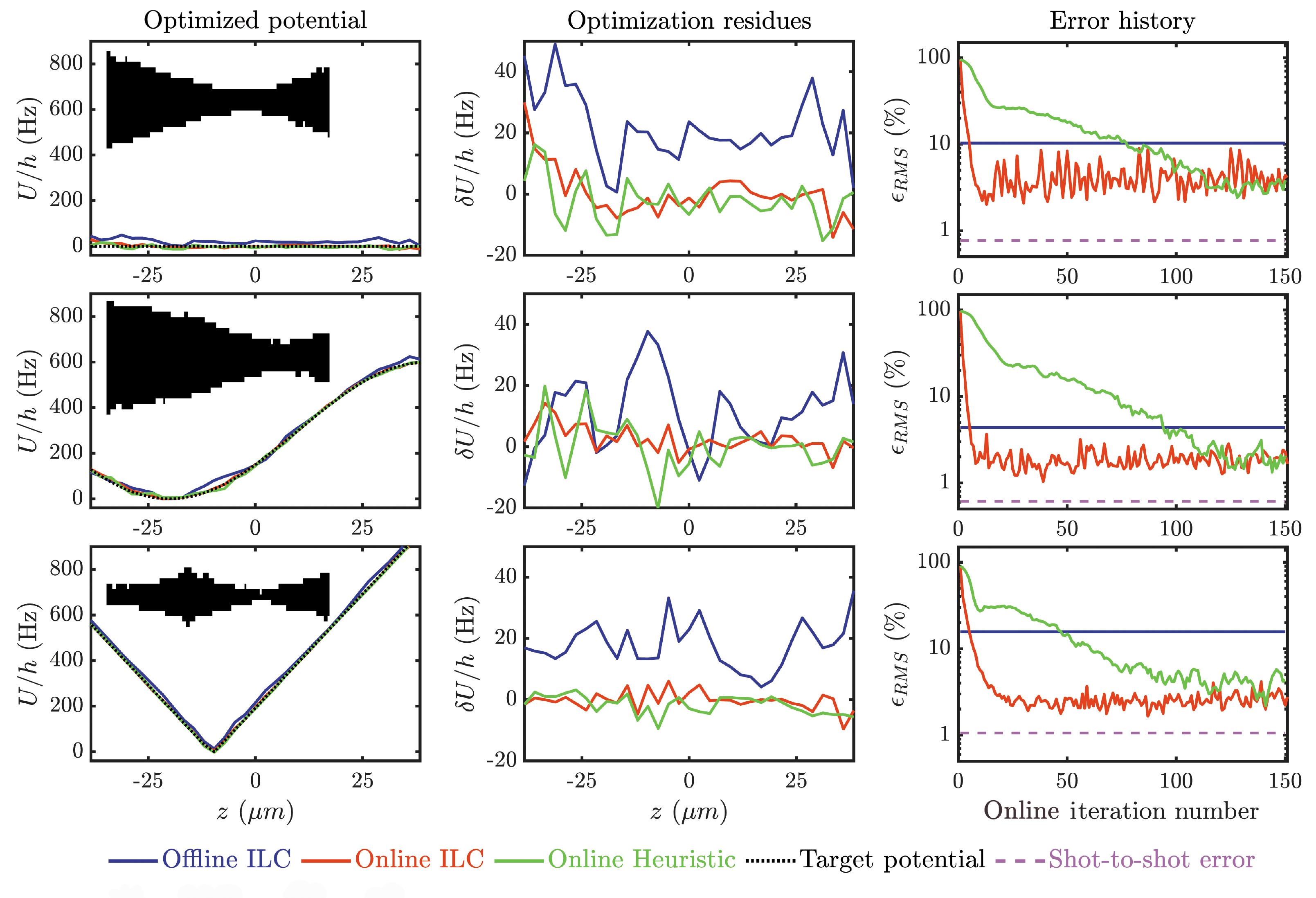}
	\caption{Optimized effective potentials assuming a 10\,Hz harmonic trap. The 1D optical potential generated by the system is inferred by reading out one selected row of the CCD. In the first column, we show the optimized effective potentials $U$ with the offline ILC, online ILC, and the heuristic online optimization (blue, orange, and green solid lines, respectively). The black dotted line shows the effective target potential for atoms (see main text for details). The insets show DMD patterns for optimization with online ILC. All three methods give a good qualitative match with the target potentials. The second column depicts the differences $\delta U$ between the target potential and the optimized potential for all 3 methods. The online ILC and the heuristic online optimization give a compatible level of optimization while the offline ILC exhibits a larger deviation from the target. In the third column, the error histories for online methods are shown in the semi-log scale, computed at each point using Eq.\eqref{eq:experr}. The blue line indicates the level of offline optimization error, which is shown as a constant reference line since it requires no online iteration, showing that the offline prediction can save 50-90 heuristic iterations. The online ILC saves more than 100 iterations compared to the heuristic method to reach a similar error level. The dashed purple line shows the shot-to-shot measurement error. 
	}
     \label{fig:opt_sled}
\end{figure*}

In this section, we introduce an optimization algorithm
based on iterative learning control (ILC) methods inspired by \cite{Deutschmann-Olek:22}.
The algorithm was used to optimize different target potentials on the experimental setup and we compared the results with the heuristic optimization method described in \cite{Tajik:19} (see Sec.\ref{sec:appendix} for details). 

\subsection{Online and Offline ILC}
ILC methods employ measurements of the considered output trajectory to iteratively solve a reference tracking problem, i.e., to find an input trajectory such that the output of a system follows a desired target trajectory as closely as possible, even in the presence of model errors and uncertainties.
The price to pay for this property is the requirement of running 
in a kind of feedback loop using experimental data.
Therefore, we show how to employ ILC algorithms either using feedback
from the physical experiment referred to as “online" ILC or from its digital counterpart, the physics-inspired model presented in Sec.\ref{sec:Model}, referred to as “offline" ILC. See Fig.\ref{fig:ML_ILC} for a schematic of the ILC algorithm. The only difference in the algorithms is that in the first case the potential $\boldsymbol{V}$ is measured, while in the second it is predicted by the model.
This way, previous (training) data is structured by the ML model while we can seamlessly improve beyond the predictive capability of the model through further online iterations.

Let us call $\boldsymbol{\nu}^n$ the virtual input at the $n$-th iteration, and $\boldsymbol{e}^n = \boldsymbol{V}^{tar} - \boldsymbol{V}^n$
the deviation from the target.
Following standard ILC approaches, the correction of the virtual input is obtained by convolution, denoted by $*$ (see Sec.\ref{sec:appendix}), with an update filter $\boldsymbol{L}_{\nu}$,
and is then added to the old virtual input to update it, i.e.,
\begin{equation}
    \boldsymbol{\nu}^{n+1} =  \boldsymbol{\nu}^{n} +  \boldsymbol{L}_{\nu} * \boldsymbol{e}^n\,.
\end{equation}
The process is repeated until either convergence or the desired error level is reached.
In order to choose an appropriate update filter, we approximate the system
in a linear and time-invariant form
\begin{equation}
    \boldsymbol{V} \approx \boldsymbol{g}_z * \boldsymbol{\nu}\,.
\end{equation}

In practice, $\boldsymbol{V}$ is recorded for a trial input,
and then fitted using a Gaussian guess for the longitudinal PSF $g_z(z) = A \exp(-(z-z_0)^2/(2\sigma^2))$,
to obtain estimates for the parameters $z_0,\sigma,A$.
We then use a pseudo-inversion-based update filter \cite{ghosh_pseudoinverse-based_2002} 
\begin{equation}
    \boldsymbol{L}_{\nu}  = \frac{\overline{\boldsymbol{G}}}{\gamma_{\nu} + \overline{\boldsymbol{G}}\boldsymbol{G}}
\end{equation}
where $\boldsymbol{G} = \mathcal{F}[\boldsymbol{g}_z]$ is the discrete Fourier transform
of the Gaussian PSF, $\overline{\boldsymbol{G}}$ its complex conjugate and all the operations are to be understood element-wise.
Here, $\gamma_{\nu} > 0$ is a regularization parameter of the system inversion which effectively reduces the high-frequency content of the input updates and, 
therefore, of the explored virtual inputs $\boldsymbol{\nu}^n$. In the presence of measurement noise, optimal choices for $\gamma_{\nu}$ are ultimately given by the experimental signal-to-noise ratio \cite{deutschmann-olek_stochastic_2021}.
The results shown in the main text are obtained using $\gamma_{\nu} = 0.1\ \max\limits_i|G_i|^2$.

\subsection{Experimental results}
We tested both the offline and online ILC algorithms by optimizing three target potentials, using the heuristic method as a reference. The results are shown in Fig.\ref{fig:opt_sled}.
In order to emulate an experiment with a trapped BEC, we represent the longitudinal magnetic trap with a harmonic potential $V_{mag}(z) = \frac{1}{2}m \omega^2 z^2$ with $\omega = 2\pi\times 10$~Hz and $m$ the mass of a $^{87}$Rb atom.
The effective potential $U(z)$ as experienced by the atoms is the sum of the magnetic trap $V_{mag}(z)$ and the optical dipole potential $V(z)$, which is estimated from measured intensity (see Sec.\ref{sec:appendix} for details).
We design targets $V^{tar}$ so that the effective potential $U(z)$ will be either constant, sinusoidal, or linear. All methods start with the DMD completely turned off.

Qualitatively all the methods are successful, as the final results are barely distinguishable from the target potential. 
For quantitative evaluation of the optimization performance, we compute the locally normalized root mean square error (RMS):
\begin{equation}\label{eq:experr}
    \epsilon_{RMS} = \sqrt{{\frac{1}{N_V}}\sum_{i=1}^{N_V} \left(\frac{V^{tar}_i - V^n_i}{V^{tar}_i}\right)^2}\,.
\end{equation}

We say that the normalization is local because it is computed point by point.
As a measure of the minimum error that can be achieved due to shot-to-shot fluctuations, we acquire 100 pictures of the optimized potential 
and compute the average $\epsilon_{RMS}$, substituting $V^{tar}$ by the average potential over the sample.

The measured optimization error of the offline ILC is shown with a constant blue line in Fig.\ref{fig:opt_sled} and ranges from 4\% to 16\% depending on the target. We find that the offline ILC scheme is able to deliver a level of accuracy already comparable to what we obtain with $\sim 100$ iterations of the heuristic algorithm, which would be equivalent to more than 1 hour of experimental time per potential shape for BEC experiments (and more than 5 hours with averaging over at least 5 images).  The optimization error for potentials obtained through offline learning is determined by the predictive capability of the ML model, which depends on the training data set. To illustrate the robustness of the physics-inspired model and its ability to extrapolate beyond training data we here train it on a generic data set consisting only of parabolic potentials.  We furthermore find that better results (with error around $3\%$) are achievable when the training data set resembles the optimized configurations more closely.

For a quasi-1D BEC with an atomic density of $100$ $\mu m^{-1}$, atomic shot noise (which limits the precision with which we can measure the potential) is at the level of 10 \% \cite{berrada_2013}, and the chemical potential is of the order of 1 kHz. Since constant offsets in the optimized potential do not have an effect on the atoms, we can neglect them. The scale of the remaining imperfections is then reduced to around 10 Hz, which is 1$\%$ of the typical chemical potential. Measuring a single potential with such precision requires averaging over 100 repetitions, which is roughly an hour of experimental time. 

We can compare these results with \cite{Tajik:19}, where the online heuristic scheme gave $\epsilon_{RMS} \sim (4 - 6)\%$ for the atom density in a similar setup. These numbers suggest that the offline ILC alone is capable of generating potentials with comparable accuracy to existing heuristic schemes. Lifting the necessity of online iterations might be a particularly appealing choice for time-dependent potentials.

The online ILC and heuristic algorithms both converge to the same error level.
Yet online ILC reaches convergence in around $\sim 10$ iterations while the heuristic algorithm needs around $\sim 100$ iterations, which is a great advantage for schemes incorporating experimental feedback. 
Moreover, the ILC algorithm does not need to manually select an optimization schedule, resulting in increased flexibility and bypassing time-consuming parameter tuning. 
Convergence time can be decreased even more by using the result of offline optimization as an initial guess for online ILC. In this case, we find that the online ILC reaches convergence in only a few iterations, resulting in an even larger speed-up compared to the heuristic method.  
In any case, the final error is far below the atomic shot noise, so it is hardly accessible in static BEC configurations.

Unlike the online and offline ILC, the heuristic algorithm does not rely on dimensional reduction and the concept of virtual input, therefore it is not restricted to the symmetric class of patterns described by Eq.\eqref{eq:VRTtoDMD}. The fact that the same error level can be achieved despite this restriction implies that this choice does not constitute a significant bottleneck for potential optimization. While other mappings are possible, such as the optimized dithered columns suggested in \cite{Deutschmann-Olek:22}, we found that symmetric mapping offers several advantages. In fact, it is easier to realize (as it does not require the solution of an additional optimization problem) and also more robust against variations of the beam shape along the transversal direction.
 
We ran the same tests using a single-frequency laser as a light source (see Sec.\ref{sec:appendix} for details).
The physics-inspired model performs worse in the prediction of laser-generated potentials, therefore the output of offline ILC is noticeably worse than the predictions for SLD. On the other hand, the online ILC algorithm optimizes the optical potentials created with the laser to the same error level as with the SLD. 
Based on these findings, we can state that our method is well suited for experimental setups employing SLDs as well as single-frequency lasers. 

%%%%%%%%%%%%%%%%%%%%%%%%%%%%%%%%%
\section{Conclusions}\label{sec:Conclutions}
%%%%%%%%%%%%%%%%%%%%%%%%%%%%%%%%%

In this paper, we presented our experimental setup for generating and efficiently optimizing 1D optical potentials. We combined a digital micromirror device for potential shaping control together with a SLD light source. We performed measurements estimating the quantitative difference between SLD's and laser's coherence properties and showing the advantages of using SLD due to its generally linear behavior.

We have implemented learning algorithms that enable 
efficient optimization of optical potentials. We have shown how to build a physics-inspired model, which acts as a digital twin of the experimental setup. The model is able to recreate the main features of the optical system based on a small set of experimental data without the need to use deep (neural) networks and large training data sets, with the advantage of saving experimental time.

The application of the Iterative Learning Control optimization method provides a more than ten-fold speed-up compared to heuristic approaches. The ILC algorithms used offline with the trained models are able to optimize optical potentials with a precision acceptable for most experiments with trapped 1D ultracold gases. Using online ILC with experimental feedback allows us to optimize optical potentials to error levels comparable to measurement error,
giving a ten-fold speed-up compared to more straightforward heuristic algorithms. By combining both ILC strategies, namely using the result of an optimized digital twin configuration as an initial guess for online ILC, we get the optimized potential with just a few experimental iterations.

Regarding optimization performance for the SLD and laser, we find that the SLD outperforms the laser using offline ILC. However, the online ILC performs equally well for laser and SLD giving the same level of optimization error. The model we developed combined with Iterative Learning Control provides a very fast way to optimize optical potentials with a DMD which might be used in a large variety of experimental setups. Our work offers a prospect for fast optimization of optical dipole potentials which is very important for time-costly experiments or for very large sequences of patterns in dynamic situations.

%%%%%%%%%%%%%%%%%%%%%%%%%%%%%%%%%
\section{Appendix} \label{sec:appendix}
%%%%%%%%%%%%%%%%%%%%%%%%%%%%%%%%%
\subsection{DMD mount}
Due to the specific construction of the Texas Instruments DLP650LNIR DMD's micromirror control mechanism \cite{Lee2008}, the DMD is mounted 45$^{\circ}$ rotated so all the optical elements are placed in one plane. Any pattern getting rotated 45$^{\circ}$ right before projecting on the DMD.
We verified that during optimization the rotation only leads to distortions of the potential which are below the resolution and therefore optimization is not affected.

\subsection{Intensity to optical dipole potential conversion}
The number of pixels in the output $V_i$ as obtained from the camera 
does not necessarily coincide with the number of pixels in the input $\nu_i$.
In order to use the model, we first interpolate $V_i$
to the input grid size, using the \texttt{interp1}
function in Matlab.

We assume the relation between light intensity
and optical dipole potential to be linear $V = \alpha_V I$
with $\alpha_V$ as found in \cite{Grimm:00}.
Since we work with red-detuned light, $\alpha_V$ is negative.
We also suppose the relation between the CCD readout $R$ and intensity to be linear.
We then compute $\alpha_{CCD} = Ir_{pow}/R $ by measuring the light intensity with a power meter. To not saturate the CCD we use a reduced amount of light intensity. To calculate the finally expected dipole potentials we employ a factor $r_{pow} = I_{full}/I_{low}$ that accounts for the source
operating at low power.

\subsection{Mathematical details}
This paper heavily relies on the discretization of functions of a real variable $f(z)$ in order to obtain finite size vectors.
If we define a coordinate grid $z_i = (i-1) \Delta - \bar{z}$ for $i=1,...,n$, then we refer to any discretized function
as $f_i = f(z_i)$ and we denote with $\boldsymbol{f}$ the $\mathbb{R}^n$ vector whose elements are $f_i$.

Let us call $\mathcal{F}$ the discrete Fourier transform acting on a vector of size $n$,
and $\mathcal{F}^{-1}$ its inverse:
\begin{align}
    \mathcal{F}[\boldsymbol{a}]_k &= \sum_{j=1}^{n} a_j e^{-\frac{2 \pi i}{n}(j-1)(k-1)} \\
    \mathcal{F}^{-1}[\boldsymbol{b}]_k &= \frac{1}{n} \sum_{j=1}^{n} b_j e^{\frac{2 \pi i}{n}(j-1)(k-1)}
\end{align}
Then, we can define the discrete convolution of two vectors $\boldsymbol{a} * \boldsymbol{b}$ as 
\begin{equation}
    \boldsymbol{a} * \boldsymbol{b} = \mathcal{F}^{-1}[\mathcal{F}[\boldsymbol{a}] \mathcal{F}[\boldsymbol{b}]]
\end{equation}
where the product on the right-hand side is element-wise.

\subsection{Heuristic algorithm}
In order to assess the advantages of the ILC methods, we employed an adapted version of the heuristic algorithm described in \cite{Tajik:19} as a reference.
It is an iterative algorithm that updates the state of each pixel based on the local differences between measured and target potentials.
The optimization happens in two phases, the first fast but rough and the second slower but more precise. During the rough phase (see Fig.\ref{fig:opt_sled} first $\sim$20 iterations) in each column DMD pixels are turned on until the difference gets lower than the chosen threshold. During the precise phase pixels can be moved away or turned off.

\subsection{Laser vs SLD}
We show in Tab. \ref{tab:lasersld} a detailed comparison of the performance of the offline and online ILC 
algorithms for the SLD and laser light sources. The values of $\epsilon_{RMS}$, cf. Eq.\eqref{eq:experr}, should be compared with the shot-to-shot errors of 0.01 for the SLD and 0.02 for the laser, which are the error components that cannot be eliminated via optimization.

\begin{table}
\begin{tabular}{|c|c|c|c|}
\hline
& Target 1& Target 2 & Target 3\\
\hline
SLD, offline & 0.10 & 0.04 & 0.16 \\
Laser, offline & 0.15 & 0.13 & 0.18 \\
\hline
SLD, online & 0.02 & 0.01 & 0.02 \\
Laser, online & 0.02 & 0.02 & 0.03 \\
\hline
\end{tabular}
\caption{ Values of $\epsilon_{RMS}$ for SLD and laser sources.}
\label{tab:lasersld}
\end{table}

\subsection*{Acknowledgments}
We would like to thank Mohammadamin Tajik and Jo\~{a}o Sabino for the discussions and technical support. This project was funded by the DFG/FWF CRC 1225 'Isoquant', Project-ID 273811115, by the Austrian Science Fund (FWF) P~36236-N (financed by the European Union - NextGenerationEU), by the Deutsche Forschungsgemeinschaft (DFG, German Research Foundation) under Germany’s Excellence Strategy – Cluster of Excellence Matter and Light for Quantum Computing (ML4Q) EXC 2004/1 – 390534769, and from the German
Federal Ministry of Education and Research through
the funding program quantum technologies—from basic
research to market under the project FermiQP,
13N15891.
M.P. has received funding from the European Union's Horizon 2020 research and innovation program under the Marie Skłodowska-Curie grant agreement No 101032523.

\bibliography{2_RefDMD}

\end{document}